\documentclass[reqno,7pt]{amsart}
\usepackage{amsmath,amssymb,epsf,eepic,epsfig}
\usepackage{mathrsfs}
\usepackage[matrix,frame,arrow]{xypic}\input{Qcircuit}
\newcommand{\cgate}[1]{*{\xy *+<.6em>{#1};p\save+LU;+RU **\dir{-}\restore\save+RU;+RD **\dir{-}\restore\save+RD;+LD **\dir{-}\restore\POS+LD;+LU **\dir{-}\endxy} \cw}

\def\qed{$\,\blacksquare$\par}
\def\vec#1{{\boldsymbol{#1}}}
\def\<{\langle}\def\>{\rangle}\def\kk{\rangle\!\rangle}\def\bb{\langle\!\langle}
\def\N#1{\left|\!\left|#1\right|\!\right|}
\def\set#1{{\sf #1}}\def\alg#1{{\mathcal #1}}\def\map#1{{\mathscr{#1}}}
\def\Bnd#1{\set{B(#1)}}\def\sH{\set{H}}\def\sS{\set{S}}\def\sE{\set{E}}
\def\Span{\set{Span}}\def\Rng{\set{Rng}}\def\Supp{\set{Supp}}\def\Spec{\set{Sp}}
\def\Reals{\mathbb R}
\def\transp#1{{#1}^\intercal}\def\dual#1{{#1}^\intercal}
\def\Tr{\operatorname{Tr}}\def\diag{\operatorname{diag}}\def\Conv{\operatorname{conv}}
\def\conv#1{{\mathscr{#1}}}\def\bP{{\mathbf P}}\def\bQ{{\mathbf Q}}\def\bM{\mathbf{M}}
\def\bR{{\mathbf R}}\def\bI{{\mathbf I}}

\def\Rip{\supset_r} 
\newtheorem{lemma}{Lemma}[section]
\newtheorem{proposition}[lemma]{Proposition}
\newtheorem{corollary}[lemma]{Corollary}
\newtheorem{definition}[lemma]{Definition}
\newtheorem{theorem}[lemma]{Theorem}
\def\remark{\medskip\par\noindent{\bf Remark. }}
\def\Proof{\medskip\par\noindent{\bf Proof. }}
\makeatletter

\def\theenumi{\@roman\c@enumi}

\def\theenumii{\@alph\c@enumii}

\def\theenumiii{\@arabic\c@enumiii}

\def\theenumiv{\@Alph\c@enumiv}
\makeatother
\begin{document}
\title{Clean Positive operator valued measures}
\author{Francesco Buscemi}
\email{buscemi@fisicavolta.unipv.it}
\address{{\em QUIT} Group, http://www.qubit.it, Dipartimento di Fisica "A. Volta", via Bassi 6,
I-27100 Pavia, Italy}
\author{Giacomo Mauro D'Ariano}
\email{dariano@unipv.it}
\address{{\em QUIT} Group, http://www.qubit.it, Dipartimento di Fisica "A. Volta", via Bassi 6,
I-27100 Pavia, Italy, and \\ Department of Electrical and Computer
Engineering, Northwestern University, Evanston, IL  60208}
\author{Michael Keyl}
\email{M.Keyl@tu-bs.de}
\address{{\em QUIT} Group, http://www.qubit.it, Dipartimento di Fisica "A. Volta", via Bassi 6,
I-27100 Pavia, Italy}
\author{Paolo Perinotti}
\email{perinotti@fisicavolta.unipv.it}
\address{{\em QUIT} Group, http://www.qubit.it, Istituto Nazionale di Fisica della Materia,
Unit\`a di Pavia, Dipartimento di Fisica "A. Volta", via Bassi 6,
I-27100 Pavia, Italy}
\author{Reinhard F. Werner}
\email{r.werner@tu-bs.de}
\address{Institut f\"ur Mathematische Physik, TU Braunschweig, Mendelssohnstr. 3 / 38106
  Braunschweig / Germany } 
\date{\today}
\maketitle
\markboth{F. BUSCEMI, G. M. D'ARIANO, M. KEYL, P. PERINOTTI, R. WERNER}{UNDISTURBED POSITIVE OPERATOR VALUED MEASURES}  
\begin{abstract}
  In quantum mechanics the statistics of the outcomes of a measuring
  apparatus is described by a positive operator valued measure (POVM).
  A quantum channel transforms POVM's into POVM's, generally
  irreversibly, thus loosing some of the information retrieved from
  the measurement. This poses the problem of {\em which POVM's are
    "undisturbed", namely they are not irreversibly connected to
    another POVM}.  We will call such POVM's {\em clean}. In a sense,
  the clean POVM's would be "perfect", since they would not have any
  additional "extrinsical" noise. Quite unexpectedly, it turns out
  that such {\em cleanness} property is largely unrelated to the
  convex structure of POVM's, and there are clean POVM's that are not
  extremal and vice-versa. In this paper we solve the cleannes
  classification problem for number $n$ of outcomes $n\leq d$ ($d$
  dimension of the Hilbert space), and we provide a a set of either
  necessary or sufficient conditions for $n>d$, along with an iff
  condition for the case of
  informationally complete POVM's for $n=d^2$.\\

\end{abstract}
\maketitle
\tableofcontents
\thispagestyle{empty}
\newpage
\section{Introduction}
The new quantum information technology\cite{pop} has resurrected the
interest in the theory of quantum measurements\cite{Busch} and
quantum open systems\cite{Davies,Kraus83a}, shifting the interest from
applications to high-sensitivity and high-precision
experiments\cite{Meystre83} to its use in quantum information
processing\cite{Nielsen2000}. Depending on the particular kind of
quantum processing---e.~g.  teleportation\cite{ben,bra}, entanglement
detection\cite{wit} and distillation\cite{ben2}---that are used in
quantum computation\cite{pop,Nielsen2000} and quantum
crypto\-graphy\cite{gisirev}, various new types of quantum
measurements are now needed. The theory for engineering new quantum
measurements optimized according to given criteria has been pioneered
since the late 60' by many authors\cite{history}
who concurred to the making of the Quantum Estimation
Theory\cite{helstrom}, the ancestor of the modern Quantum Information
Theory.

The general strategy of quantum estimation theory is to optimize the
output statistics of the measuring apparatus according to a given
criterion/fidelity, which depends on the specific use of the
measurement, the outcome statistics of the measurement for all
possible input states being described by a positive operator valued
measure (POVM)\cite{helstrom}. POVM's form a convex set, where convex
combinations correspond to random choices among different apparatuses.
Most optimization problems actually resort to minimize a concave
function on such convex set, thereby optimization can be restricted to
its extremal points, where concave functions attain their minimum.
Coincidentally, due to the specific form of the optimization function,
in many applications the optimal POVM's turn out to have unit
rank---e.~g. for phase estimation on pure
states\cite{helstrom,holevo}---and this has led to the widespread
belief that optimality is synonym of rank-one, whereas for
sufficiently large dimension, and typically for optimization with
input mixed states, the rank of extremal POVM's can be easily larger
than one, as shown in Refs. \cite{extrpovm,cov_sieve,POVMext}.

\begin{figure}[htb]\label{f:prepost}
\begin{alignat*}{2}
  &\Qcircuit @C=1em @R=.7em @! R {&\qw &\meter} \qquad&\Qcircuit
  @C=1em @R=.7em @! R { &\equiv &}&\qquad \Qcircuit @C=1em @R=.7em @!
  R {&\qw &\qw & \meter & \cw & \cgate{\mbox{data processing}}
    \gategroup{1}{3}{1}{6}{2.2em}{.} }\tag{A}\\
  &\Qcircuit @C=1em @R=.7em @! R {&\qw &\meter} \qquad&\Qcircuit
  @C=1em @R=.7em @! R { & \equiv &}&\qquad \Qcircuit @C=1em @R=.7em @!
  R {&\qw &\qw &\gate{\map{E}} &\meter
    \gategroup{1}{3}{1}{5}{2.2em}{.} }\tag{B}
\end{alignat*}
\caption{There are two ways of processing POVM's: (A) the {\em
    post-processing} of the output data, and (B) {\em pre-processing}
  of the input state by a quantum channel.  The post-processing cannot
  generally achieve the same result of a {\em pre-processing}: the
  post-processing is purely classical, whereas the pre-processing is
  quantum.}
\end{figure}
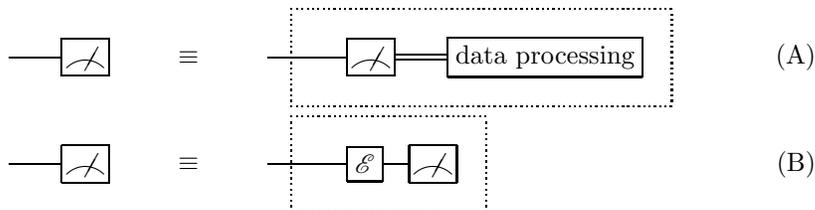

In a specific application the optimal POVM does not necessarily attain
the whole accessible information. At first sight, this assertion may
appear contradictory: how a POVM can be optimal, if it wastes
accessible information?  However, once the measurement is performed,
no other possibility for optimization is left apart from the
processing of the outcome---{\em post-processing} for short---and,
being purely classical, the post-processing cannot generally achieve
the same result of a {\em pre-processing} by a quantum channel.  The
situation is depicted in Fig. \ref{f:prepost}.
Clearly, the pre-processing can change the POVM irreversibly,
reducing the information from the measurement.  On the other hand, it
is possible that a POVM optimal for a given criterion is obtainable
from another {\em cleaner} one via an irreversible pre-processing as
in Fig. \ref{f:prepost}B. This means that in some cases we need to
give up some {\em quantity} of information for the {\em quality} of
the information.

The above scenario poses the problem of {\em which POVM's are
  "undisturbed"}, namely {\em are not irreversibly connected to
  another POVM}. We will call such POVM's {\em clean}---in a sense a
clean POVM would be "perfect", since it would not have any additional
"extrinsical" noise, or it has lost no information irreversibly. Quite
surprisingly, as announced, in this paper we will see that the {\em
  cleanness} property of the POVM is largely unrelated to its
extremality, and there are clean POVM's that are not extremal and
vice-versa. The problem of classifying clean POVM's turns out to be
more difficult than that of classifying extremal ones, and in this
paper we will give a complete classification of clean POVM's only for
a number $n$ of outcomes $n\leq d$, whereas for $n>d$ we will give a
set of interesting necessary conditions, and an iff condition for the
case of informationally complete POVM's for $n=d^2$. Clearly, the need
for a number of outcomes $n>d$ can be required by the particular
optimization problem (see, for example, Refs. \cite{Shor,Fuchs}),
however, no more than $n=d^2$ elements are needed, which is the
maximum number of outcomes for extremality\cite{extrpovm}. Davies
\cite{DaviesIEE} proved $d^2$ to be an upper bound for the maximal
cardinality of the POVM needed to attain the accessible information,
and still it is debated if $d^2$ outcomes are actually needed (the
cases of Refs. \cite{Shor,Fuchs} proved that the lower bound is
actually larger than $d$).  This difficulties reflect those in
classifying cleanness for $n>d$.  In a sense it is clear that $d^2$
elements are needed to retrieve the accessible information, when the
kind of information needs to be decided after the measurement has been
performed. Indeed, an extremal POVM with $d^2$ outcomes is versatile
to any kind of information encoding, as it is "informationally
complete"\cite{univest}, namely it makes possible to estimate any
ensemble average by changing only the data-processing of the outcomes
(such an extremal measurement with $d^2$ elements can be proved to
exist for any dimension $d$\cite{extrpovm}). Clearly, for an extremal
informationally complete measurement, a further optimization step can
be achieved at the level of data processing\cite{infogroup,bellmeas},
once the kind of information of interest has been decided.  Thus, the
post-processing of the measurement is still a useful tool in
retrieving the right information from a measurement.
 
The paper is organized as follows. After introducing some notations
and prerequisites in section \ref{s:notation}, in section
\ref{s:usefl} we state some general results about channels and POVM's
which will be used throughout the paper. In section \ref{s:chanPOVM}
we analyze the convex set of channels connecting two POVM's.  Section
\ref{s:post} is devoted to a complete analysis of the classical
post-processing, and give a complete characterization of "cleanness"
under post-processing.  Section \ref{s:pre} addresses the problem of
the pre-processing ordering of POVM's, namely if a POVM is "cleaner"
than another, and when they are "equivalent", which corresponds to the
possibility of reversing the action of the channel connecting the two
POVM's. Section \ref{s:pre2} shows that for dimension $d=2$
equivalence under cleanness is the same as unitary equivalence.
Section \ref{s:pre-under} fully solves the case of number of outcomes
$n\leq d$, and gives some interesting alternative theorems for the
case of {\em effects}, namely the two-outcome POVM's.  Sect.
\ref{s:preinfo} analyzes the case of informationally complete POVM's,
giving also a iff condition characterizing the clean POVM's. Section
\ref{s:pre-rank} gives some conditions for rank-one measurements.
Finally, we conclude the paper in Sect. \ref{s:conclusion} with a list
of most relevant results and of the main open problems.

\section{Notation and pre-requisites}\label{s:notation}
Throughout this paper we will consider a quantum system with Hilbert
space $\sH$ with finite dimension $d=\dim(\sH)$, and denote by $\sS$
the set of states on $\sH$ (corresponding to positive unit-trace
operator on $\sH$), and by $\Bnd{H}$ the algebra of bounded operators
on $\sH$. We will use capital script fonts e.~g.
$\alg{A},\alg{B},\ldots$ to denote operator algebras in $\Bnd{\sH}$,
and with the symbol $\alg{A}'$ we will denote the commutant of
$\alg{A}$, namely the algebra defined as
$\alg{A}'\doteq\{Y\in\Bnd{\sH}|[X,Y]=0,\;X\in\alg{A}\}$.  Completely
positive trace-preserving (CPT) and identity-preserving (CPI) maps on
$\sS$ and $\Bnd{H}$, respectively---all generally referred to as {\em
  channels}---will be denoted by capital calligraphic letters, e.~g.
$\map{A},\map{B},\ldots$, whereas we will always use capital Roman
letters for operators. We will restrict attention to POVM's
$\{P_e\}_{e\in\sE}$ with finite sampling space $\sE$, namely
\begin{equation}
P_e\geq 0,\;\forall e\in\sE,\quad\sum_{e\in\sE}P_e=I.
\end{equation}
We will use extensively the vector notation $\bP\equiv\{P_e\}$,
$\sE(\bP)$ denoting the sampling space of $\bP$, and $|\bP|$ the
cardinality of $\sE(\bP)$. The vector notation will be naturally
extended to tensor products---e.~g.  $\bP\otimes\bQ$ for the POVM
$\{P_e\otimes Q_f\}_{e\in\sE(\bP),f\in\sE(\bQ)}$ on
$\sH\otimes\sH$---and to functionals---e.~g. $\Tr[\rho\bP]$ for the
vector of probabilities $\Tr[\rho P_e]$. By $\Span(\bP)$ we will
denote the linear operator space spanned by the POVM elements
$\{P_e\}_{e\in\sE(\bP)}$, and by $\Rng(\bP)$ the range of the POVM
$\bP$, which is defined as the following convex subset of
$\Reals_+^{|\bP|}$
\begin{equation} \label{eq:1}
\Rng(\bP)\doteq\left\{\Reals_+^{|\bP|}\ni\vec p=\Tr[\rho\bP],\;\rho\in\sS\right\}.
\end{equation}
The convex set of POVM's with cardinality $N$ will be denoted by
${\conv{P}_N}$.\par

Finally, we will use the symbol $|A\kk$ to denote the following bipartite
vector in $\sH\otimes\sH$
\begin{equation}
|A\kk\doteq\sum_{m,n=1}^{d}A_{m,n}|m\>|n\>\,,
\end{equation}
where $A\in\Bnd\sH$ is the operator corresponding to the $d\times d$
matrix with elements $A_{m,n}$ on the basis $\{|n\>\}$. One can easily
verify the following useful identities
\begin{align}
&A\otimes \transp B|C\kk=|ACB\kk\nonumber\\
&\Tr_1[|A\kk\bb B|]=\transp A B^*\\
&\Tr_2[|A\kk\bb B|]=A B^\dag\nonumber\,,
\end{align}
where $\transp X$ denotes the transpose in the basis $\{|n\>\}$, while
$X^*$ is the complex conjugate in the same basis. $\Tr_i$ denotes the
partial trace on the $i$-th space.\par 

\section{Useful lemmas about channels and POVM's}\label{s:usefl}

In the following we will name a map $\map{E}$ {\em
  spectrum-width decreasing} when it reduces the "spectral width" of a
real symmetric operator $X$, namely when 
\begin{equation}
[\lambda_m(\map{E}(X)),\lambda_M(\map{E}(X))]\subseteq [\lambda_m(X),\lambda_M(X)],
\end{equation}
$\lambda_M(X)$ and $\lambda_m(X)$ denoting the maximum and minimum
eigenvalues of $X$, respectively.
\begin{lemma}\label{lem:eigmap} Channels are spectrum-width decreasing.
\end{lemma}
\Proof Consider the eigenvector $|\psi_j\>$ of $\map{E}(X)$
corresponding to the eigenvalue $\lambda_j(\map{E}(X))$. One has
\begin{equation}
\lambda_j(\map{E}(X))=\Tr[\map{E}(X)|\psi_j\>\<\psi_j|]=\Tr[X\dual{\map E}(|\psi_j\>\<\psi_j|)]
\in[\lambda_m(X),\lambda_M(X)],
\end{equation} 
since the dual map $\dual{\map E}$ is CPT. \qed \medskip

Notice that in the above lemma the identity-preserving condition is
crucial, since the lemma would not hold for a CPT map $\map{E}$, e.~g.
$\map{E}(\rho)=|\psi\>\<\psi|$, and the spectral width increases from
$[\lambda_m(\rho),\lambda_M(\rho)]$ to $[0,1]$.

The inverse of a non-unitary invertible channel is necessarily not
completely positive.
\begin{theorem}[Wigner]\label{th:wigner}
  Any invertible channel has CP inverse iff it is unitary.
\end{theorem}
\Proof Let $\map{E}_1$ and $\map{E}_2$ be two channels such that
$\dual{\map{E}_2}\circ \dual{\map{E}_1}(\rho)=\rho$. Hence:
\begin{equation}\label{Kraus-comp}
  \sum_{ij}B_jA_i|\psi\>\<\psi|A_i^\dag B_j^\dag=|\psi\>\<\psi|,\qquad\forall|\psi\>,
\end{equation}
where $A_i$ and $B_j$ are canonical Kraus representations for
$\map{E}_1$ and $\map{E}_2$, respectively.  Since all terms in the sum
are positive, this means that
$B_jA_i|\psi\>=\beta_{ij}^{\psi}|\psi\>$, for all $|\psi\>$ and all
$i,j$. By linearity, it is clear that $\beta_{ij}$ cannot depend on
$|\psi\>$, implying that $B_jA_i=\beta_{ij}I$, for all $i,j$.\par

We can now prove that $\beta_{ij}\neq 0$, for all $i,j$. Otherwise,
there exists a couple of operators $B_k$ and $A_l$ for which
$B_kA_l=0$. These two operators must both be non invertible, since, if
one is invertible, the other has to be null, and we can w.l.o.g.  drop
it from the Kraus representation (\ref{Kraus-comp}). Let us fix the
couple $k,l$ for which $B_kA_l=0$, namely both are not invertible.
Now, the only possibility to have $B_jA_i=\beta_{ij}I$ for all $i,j$
is that $B_kA_i=0$ for all $i$ (since $B_k$ is not invertible, whence
necessarily $B_kA_i$ cannot be full-rank), and analogously $B_jA_l=0$
for all $j$. In this case, all $B_j$'s supports would be forced to be
contained in the orthogonal complement to the range of $A_l$ (which is
strictly contained in the full Hilbert space), and this would be in
contradiction with the normalization condition $\sum_jB_j^\dag B_j=I$.
Therefore, $\beta_{ij}\neq 0$ for all $i,j$, and the operators $A_i$
and $B_j$ are all invertible. This allows us to write
\begin{equation}
\begin{split}
B_j&=\beta_{ij}A_i^{-1},\qquad\forall j,\\
A_i&=\beta_{ij}B_j^{-1},\qquad\forall i,
\end{split}
\end{equation}
whence all $B_j$'s are proportional to each other, and analogously for the $A_i$.
In other words, the Kraus representations of $\map{E}_1$ and
$\map{E}_2$ are made of only one operator.  This means that
$\map{E}_1$ and $\map{E}_2$ are unitary, one the inverse of the other.

The converse direction is trivial. In Corollary \ref{l:invneg}, we
will prove that the inverse map of an invertible non-unitary channel
is indeed non-positive. \qed \medskip

\begin{theorem}[Chefles, Jozsa, Winter]\label{th:caves}
  Consider two sets of pure states on $\sH$ with the same cardinality.
  There exist two channels mapping the elements of the first set to
  the corresponding elements of the second set and vice versa, iff the
  two sets of states are unitarily equivalent.
\end{theorem}
\Proof See Ref. \cite{chejowi}. \qed \medskip

\begin{lemma}[Lindblad]\label{lem:lind}
A channel $\map{E}$ stabilizes an algebra $\alg{S}\subseteq\Bnd{\sH}$, namely
\begin{equation}
\map{E}(X)=X,\quad\forall X\in\alg{S},
\end{equation}
iff the operators $\{E_k\}$ of any Kraus form $\map{E}(X)=\sum_kE_k^\dag XE_k$
belong to the commutant $\alg{S}'$ of the algebra $\alg{S}$.
\end{lemma}
\Proof See Ref. \cite{lindblad}. \qed
\medskip

Finally let us state some results about extendiblity of completely
positive maps (mostly taken from \cite{paulsen}). To this end let us
consider a linear subset $\alg{S}$ of $\Bnd{\sH}$ which contains the
identity and is closed under adjoints -- each set $\alg{S}$ of this
type will be called in the following an \emph{operator system}. It is
easy to see that $\alg{S}$ is generated (as a linear space) by its
positive elements. It makes therefore sense to speak about positive
maps $\map{E}: \alg{S} \to \alg{A}$ into an algebra $\alg{A}$ and we
can define also \emph{complete positivity} in the usual way. Now the
question arises whether such an $\map{E}$ can be extended \emph{as a
  completely positive map} to $\Bnd{\sH}$. The following theorem gives
a positive answer \cite[Thm. 6.2 and 7.5]{paulsen}:

\begin{theorem}[Arveson's extension theorem]\label{thm:arv}
  Each completely positive map $\map{E} : \alg{S} \to
    \Bnd{\sH}$ defined on an operator system $\alg{S} \subset
  \Bnd{\sH}$ can be extended to a completely positive map on
  $\Bnd{\sH}$.
\end{theorem}

If $\map{E}$ is only positive (and not necessarily completely
positive) a similar result is not available (cf. the corresponding
discussion in Section \ref{sec:pre-proc-other}). An important
exception arises however, if the algebra $\alg{A}$ is abelian
\cite[3.9]{paulsen}

\begin{theorem} \label{thm:abel-pos}
  If $\map{E}: \alg{S} \to \alg{A}$ is positive, $\alg{S}$ an operator
  system and $\alg{A}$ an abelian algebra, the map $\map{E}$ is
  completely positive. 
\end{theorem}

\section{The convex set of channels connecting two POVM's}\label{s:chanPOVM}

We now analyze the convex set of channels connecting two given POVM's
$\bP$ and $\bQ$, in equations
\begin{equation}
\conv{C}_{\bP\bQ}=\{\map{E}\,\text{channel}\;|\map{E}(\bP)=\bQ\}.
\end{equation}
The extremal elements of $\conv{C}_{\bP\bQ}$ can be characterized in
terms of the operators $\{E_i\}$ of the canonical Krauss decomposition
by the following theorem.
\begin{theorem}
  The map $\map{E}\in\conv{C}_{\bP\bQ}$ is extremal iff for some
  element $P_k$ of the POVM $\bP$ the operators
  $\{E^\dag_iP_kE_j\}_{ij}$ made with the canonical Kraus operators
  $\{E_i\}$ of the map are linearly independent.
\end{theorem}
\Proof First we show by contradiction that the condition is
sufficient. In fact, suppose that $\map{E}$, with
$\{E^\dag_iP_kE_j\}_{ij}$ linearly independent, is not extremal in
$\conv{C}_{\bP\bQ}$. Then there exist two different channels
$\map{E}_\pm\in\conv{C}_{\bP\bQ}$ such that
\begin{equation}
\map{E}=\frac12(\map{E}_++\map{E}_-).
\end{equation}
Upon defining $\map{P}\equiv\map{E}_+-\map{E}$, clearly one has
$\map{P}(\bP)=0$ and $\map{E}\pm\map{P}=\map{E}_\pm$, which are
channels. Then $R_{{\map E}_\pm}\equiv R_{\map E}\pm R_{\map P}\geq0$, where for any channel
$\map{E}$ the positive operator $R_{\map E}$ in linear correspondence with $\map{E}$ 
is defined as $R_{\map E}=\sum_{j}|E_j\kk\bb E_j|$ for $\{E_j\}$ Kraus operators of $\map{E}$
\cite{choijam}.
This implies that $\Supp(R_{\map P})\subseteq\Supp(R_{\map E})$,
namely there exists a nonvanishing matrix $p_{ij}$ such that $R_{\map P}=\sum_{ij}p_{ij}|E_i\kk\bb
E_j|$. As a consequence we have
\begin{equation}
\map{P}(P_k)=\sum_{ij}p_{ij}E^\dag_iP_kE_j=0\,,\quad\forall k\,.
\end{equation}
This contradicts the hypothesis. The proof that it is also necessary
is now straightforward. Suppose indeed that the operators
$\{E^\dag_iP_kE_j\}_{ij}$ are linearly dependent.  Then there exists a
non vanishing matrix of coefficients $a_{ij}$ such that
$\sum_{ij}a_{ij}E^\dag_iP_kE_j=0$ for all $k$. If we define
$p_{ij}=\kappa(a_{ij}+a^*_{ij})$, then the map
$\map{P}(X)=\sum_{ij}p_{ij} E_j^\dag X E_i$ will annihilate all
elements of the POVM $\bP$, namely $\map{P}(\bP)=0$. Moreover, for a
sufficiently small $\kappa\neq0$ both maps $\map{E}_\pm=
\map{E}\pm\map{P}$ will be channels and will belong to
$\conv{C}_{\bP\bQ}$. This implies that
$\map{E}=\frac12(\map{E}_++\map{E}_-)$, namely $\map{E}$ is not
extremal. \qed \bigskip

One can prove that either any element of the border of
$\conv{C}_{\bP\bQ}$ is also an element of the border of the full
convex set of channels, or $\conv{C}_{\bP\bQ}\equiv\{\map E\}$. This
comes from the definition of the border of a convex set
\begin{definition}
For a convex set $\conv{C}$, an element $p\in\conv{C}$ belongs to its boundary $\partial\conv{C}$
if and only if there exists $q\in\conv{C}$ such that 
\begin{equation}
p+\epsilon(q-p)\in\conv{C}\,,\quad
p-\epsilon(q-p)\not\in\conv{C}\,,\quad\forall\epsilon\in[0,1], . 
\end{equation}
or, equivalently iff there exists $q\in\conv{C}$ such that for all $\epsilon>0$ for which
$p+\epsilon q\in\conv{C}$ then $p-\epsilon q\not\in\conv{C}$.
\end{definition}
We will now prove the following lemma.
\begin{lemma}\label{lem:bortpq}
The border of the convex $\conv{C}_{\bP\bQ}$ is a subset of the border of the convex of all
channels.
\end{lemma}
\Proof Consider a channel $\map{E}\in\conv{C}_{\bP\bQ}$ and a ``perturbation'' $\map{P}$ such that 
$\map{E}+\epsilon\map{P}\in\conv{C}_{\bP\bQ}$ $\forall\epsilon\in[0,1]$. By definition $\map{P}(P_i)=0$
for all $P_i$, whence, if $\map{E}-\epsilon\map{P}$ is completely positive, then it necessarily
belongs to $\conv{C}_{\bP\bQ}$. Therefore, the only way to have $\map{E}$ on the border of
$\conv{C}_{\bP\bQ}$ is to have $\map{E}-\epsilon\map{P}$ not CP, namely $\map{E}$ lies on the border
of the convex of all channels.\qed 
\medskip 
A ``geometrical'' proof is also the following. Since the constraint defining $\conv{C}_{\bP\bQ}$ is
linear, then $\conv{C}_{\bP\bQ}$ is a linear section of the convex of all channels, whence its
border belongs to the border of the convex of all channels.
\remark Notice that the convex set $\conv{C}_{\bI\bI}$ will coincide with that of all
channels, $\bI=\{I\}$ denoting the trivial POVM.
\medskip
\remark
From Lemma \ref{lem:bortpq} it follows that when two POVM's are connected by a channel they can be
always connected by a border channel, apart from the case in which the connecting channel is unique.

\section{Post-processing}\label{s:post}

The most general post-processing of a POVM, is a shuffling of the POVM
elements with conditional probability $p(i|j)$, corresponding to the
mapping
\begin{equation}
Q_i=\sum_jp(i|j)P_j.
\label{eq:postproc}
\end{equation}
When two POVM's $\bP$ and $\bQ$ are connected by a mapping of the form
\eqref{eq:postproc} for some conditional probability $p(i|j)$ we will
write $\bP\succ_p\bQ$, and say that the POVM $\bP$ is {\em cleaner
  under post-processing}---{\em post-processing cleaner}, for
short---than the POVM $\bQ$.  Notice that a relation of the form
(\ref{eq:postproc}) is meaningful generally for $|\bP|\neq|\bQ|$, with
the number of outcomes changing from input to output.
\par Relevant examples of post processing are:
\begin{enumerate}
\item identification of two outcomes, e.~g. $j$ and $k$ are identified
  with the same outcome $l$, corresponding to $p(n|j)=\delta_{ln}$,
  $p(n|k)=\delta_{ln}$;
\item permutation $\pi$ of outcomes, corresponding to
  $p(\pi(j)|k)=\delta_{jk}$.
\end{enumerate}
\medskip
\par The relation $\succ_p $ is a pseudo-ordering, since it is
\begin{enumerate}
\item reflexive, corresponding to
\begin{equation}
\bP\succ_p\bP,\qquad p(i|j)=\delta_{ij};
\end{equation}
\item transitive, i.~e. $\bP\succ_p\bQ\succ_p\bR$, corresponding to
\begin{equation}
\begin{split}
  R_i=\sum_jp(i|k)Q_k,\;Q_k=&\sum_jp'(k|j)P_j,\Longrightarrow
  R_i=\sum_jp''(i|j)P_j,\\
  p''(i|j)&=\sum_kp(i|k)p'(k|j).
\end{split}
\end{equation}
\end{enumerate}
\medskip

An equivalence relation under post-processing can be defined as
follows.
\begin{definition}
  The POVM's $\bP$ and $\bQ$ are {\em post-processing equivalent}---in
  symbols $\bP\simeq_p \bQ$---iff both relations $\bP\succ_p\bQ$ and
  $\bQ\succ_p\bP$ hold.
\end{definition}
We are now in position to define {\em cleanness under post
  processing}, namely
\begin{definition}
  A POVM $\bP$ is {\em post-processing clean} if for any POVM $\bQ$
  such that $\bQ\succ_p\bP$, then also $\bP\succ_p\bQ$ holds, namely
  $\bP\simeq_p\bQ$.
\end{definition}
The characterization of cleanness under post-processing (classical) is
much easier than that of cleanness under pre-processing (quantum), and
is given by the following theorem.
\begin{theorem}
A POVM $\bP$   is post-processing clean iff it is rank-one.
\end{theorem}
\Proof First notice that a POVM $\bP$ with elements having rank higher
than one are not post-processing clean. In fact, in this case one can
diagonalize all the POVM elements and consider the POVM $\bP'$ made of
rank-one projections over all eigenvectors multiplied by the
corresponding eigenvalue. Then, clearly $\bP'\succ_p\bP$ by
identification of outcomes. In equations
\begin{equation}
P_i=\sum_{k_i}|\lambda^{(i)}_{k_i}\>\<\lambda^{(i)}_{k_i}|,\qquad
P_{i,k}'=|\lambda^{(i)}_k\>\<\lambda^{(i)}_k|,\quad
\Longrightarrow\quad\bP'\succ_p\bP,
\end{equation}
corresponding to the identification of outcomes 
\begin{equation}
p(i|j,k_j)=\delta_{ij}\;\forall k_j.
\end{equation}
Reversely, all rank-one POVM's are post-processing clean, namely if
$\bQ\succ_p\bP$, then also $\bP\simeq_p\bQ$ must hold. In fact,
suppose that $\bP$ is rank-one and that there exists a POVM $\bQ$ such
that $\bQ\succ_p\bP$, namely
\begin{equation}
P_i=\sum_{j}p(i|j)Q_j.
\end{equation}
Now, since all elements $P_i$ are rank-one, the elements $Q_j$ are
necessarily proportional to $P_i$ for all the indices $j$ such that
$p(i|j)\neq0$, namely also $\bQ$ is rank-one, with
\begin{equation}
p(i|j)Q_j=\alpha_j P_i\,,
\end{equation} 
with $\sum_j \alpha_j=1$, and $\alpha_j\geq0$. For a fixed $j$,
$p(i|j)=0$ for at least one $i$, otherwise all the $P_i$'s would be
proportional. For the same reason, for a fixed $i$, $p(i|j)=0$ for at
least one $j$. We can then collect the indices $i$ such that
$p(i|j)\neq0$ in the set $I(j)$, and write
\begin{equation}\label{QPalpha}
Q_j=\sum_ip(i|j)Q_j=\sum_{i\in I(j)}p(i|j)Q_j=\sum_{i\in I(j)}\alpha_j P_i\,.
\end{equation}
Finally, it is immediately verified that
\begin{equation}
q(j|i)=\left\{
\begin{split}
&\alpha_j,\ i\in I(j)\\
&0,\ \textrm{otherwise}
\end{split}\right.
\end{equation}
is a conditional probability since for all $i$ one has
$\sum_jq(j|i)=\sum_j\alpha_j=1$. Therefore, from Eq. (\ref{QPalpha}) it follows that we have also
$\bP\succ_p\bQ$, namely $\bP\simeq_p\bQ$.\qed

\section{Pre-processing: ordering and equivalence of POVM's}\label{s:pre}

The action of channels allows to define the following pseudo-ordering.
\begin{definition} \label{def:1}
  Given the POVM's $\bP$ and $\bQ$ with $|\bP|=|\bQ|$ we define
  $\bP\succ\bQ$ iff there exists a channel $\map{E}$ such that
\begin{equation}
\bQ=\map{E}(\bP),
\end{equation}
and we will say that the POVM $\bP$ is {\em cleaner} than the POVM
$\bQ$.
\end{definition}
\medskip


\begin{definition}
We call a POVM $\bP$ "clean" iff for any POVM $\bQ$ such that $\bQ\succ\bP$ one also has $\bP\succ\bQ$.
\end{definition}
\par It is easily proved that the
relation $\succ$ is transitive and reflexive, namely it is a
pseudo-ordering.  Let us now define the following relation
\begin{definition} We say that the two POVM's $\bP$ and $\bQ$ are equivalent---denoted as
  $\bP\simeq\bQ$---when one has both $\bP\succ\bQ$ and $\bQ\succ\bP$.
\end{definition}
Clearly $\simeq$ is an equivalence relation. The pseudo-ordering $\succ$ now defines a partial
ordering between equivalence classes. Indeed define the ordering between classes 
as follows
\begin{equation}
[\bP]\geq[\bQ]\qquad\text{iff}\qquad
\bP'\succ\bQ'\,,\quad\forall\bP'\in[\bP],\,\bQ'\in[\bQ]\,.\label{parorder}
\end{equation}
The above ordering is consistently defined, since $\bP',\bP''\in[\bP]$ means both $\bP'\succ\bP''$
and $\bP''\succ\bP'$, whence, by transitivity $\bP''\succ\bP'\succ\bQ'\succ\bQ''$, and the ordering
doesn't depend on the chosen elements of the equivalence classes. 
This proves the consistency of the definition of $\geq$. Therefore, in the following we can consider
a single element $\bP$ instead of the class $[\bP]$. In this way we can easily prove reflexivity
$[\bP]\geq[\bP]$, since $\bP\succ\bP$, and transitivity 
\begin{equation}
[\bP]\geq[\bQ],\quad[\bQ]\geq[\bR]\,\Rightarrow[\bP]\geq[\bR]\,,
\end{equation}
since $\bP\succ\bQ $, $\bQ \succ{\mathbf R}$
implies $\bP \succ{\mathbf R}$ by transitivity of $\succ$. Now
let us consider the case when both $[\bP ]\geq[\bQ ]$
and $[\bQ ]\geq[\bP ]$. Then we have $\bP\succ\bQ $ and $\bQ \succ\bP $, namely
$[\bP ]\equiv[\bQ ]$. \qed 
\bigskip 

One would be tempted to conjecture that the relation $\simeq$ is
equivalent to unitary equivalence, which is defined through
\begin{definition}
  The POVM's $\bP $ and $\bQ $ are unitarily equivalent,
  $\bP\simeq_U\bQ $ for short, iff there exists a unitary operator $U$
  such that $\bQ=U\bP U^\dag$.
\end{definition}
However, as we will see in the following, there exist equivalent
POVM's which are not unitarily equivalent.\par

We have now the following necessary condition for equivalence under
pre-processing
\begin{theorem}\label{th:maxminei}
  If $\bP\simeq\bQ$ then for each event $e\in\sE(\bP)$ we have
\begin{equation}
\lambda_M(P_e)=\lambda_M(Q_e)\equiv\lambda_M(e)\,,\quad\lambda_m(P_e)=\lambda_m(Q_e)\equiv\lambda_m(e)\,.
\end{equation}
\end{theorem}
\Proof By Lemma \ref{lem:eigmap} we have both $\lambda_M(P_i)\geq\lambda_M(Q_i)$ and
$\lambda_M(P_i)\leq\lambda_M(Q_i)$, and similarly for the minimum eigenvalues. \qed

\section{Pre-Processing: positive maps and related theorems}
\label{sec:pre-proc-other}

There are two interesting variants of the relation $\succ$ just
introduced, which help to get a more geometric insight into the
structure. The first arises, if we replace the completely positive map
$\map{E}$ in Definition \ref{def:1} by positive (but not necessarily
\emph{completely} positive) one. Hence we can define for two POVMs
$\bP$, $\bQ$ with $|\bP| = |\bQ|$ 
\begin{equation} \label{eq:3}
  \bP \gg \bQ \quad \Leftrightarrow \quad \bQ = \map{E}(\bP),\ \text{$\map{E}$ positive.}
\end{equation}
It is obvious that $\bP \succ \bQ$ implies $\bP \gg \bQ$ but the other
way round does not hold. This can be seen, if we consider an
informationally complete POVM $\bP$ and define $\bQ = \Theta(\bP)$,
where $\Theta$ denotes the transposition map (i.e. $\Theta(A) =
\transp A$).  Positivity of $\Theta$ implies $\bP \gg \bQ$. But
$\Theta$ is only positive and not completely positive and it is the
only map which connects $\bP$ and $\bQ$. The latter follows from
informational completeness of $\bP$, because this implies that the
elements of $\bP$ are a basis of $\Bnd{\sH}$. Hence $\bP \succ \bQ$
\emph{does not hold}.

Now consider the \emph{ranges} $\Rng(\bP)$, $\Rng(\bQ)$ of $\bP$ and
$\bQ$, defined in Equation (\ref{eq:1}). If $\vec{p} \in \Rng(\bQ)$
there is by definition a density operator $\rho$ with $\vec{p} =
\Tr[\bQ \rho]$. Hence, $\bP \gg \bQ$ implies
\begin{equation}
  \vec{p} = \Tr[\bQ \rho] = \Tr\bigl[\map{E}(\bP) \rho\bigr] = \Tr\bigl[\bP
  \dual{\map{E}}(\rho)\bigr] 
\end{equation}
and therefore we get $\vec{p} \in \Rng(\bP)$. This observation motivates the
definition:
\begin{equation}\label{martellata}
  \bP \Rip \bQ \quad \Leftrightarrow \quad \Rng(\bQ) \subset \Rng(\bP).
\end{equation}
According to our previous discussion we get in this way a hierarchy of
relations
\begin{equation}
  \bP \succ \bQ \Rightarrow \bP \gg \bQ \Rightarrow \bP \Rip \bQ.
\end{equation}
We have already seen that the direction of the implication between $\succ$
and $\gg$ can not be reversed. For $\gg$ and $\Rip$ this is more
difficult. To see that they are (very) closely related, consider the
linear hull $\Span(\bP)$ of the elements of $\bP$, which is obviously
an \emph{operator system} (cf. Section  \ref{s:usefl}). Hence we can
speak about positive linear maps from $\Span(\bP)$ to $\Span(\bQ)$. This
fact can be used to characterize the relation $\Rip$ in the following
way:

\begin{proposition} \label{prop:1}
  Consider two POVMs $\bP$, $\bQ$ with $|\bP| = |\bQ|$. Then the
  following statements are equivalent:
  \begin{enumerate}
  \item \label{item:1} 
    $\bP \Rip \bQ$
  \item \label{item:2}
    There is a (unique) positive map $\map{E}: \Span(\bP) \to
    \Span(\bQ)$ with $\map{E}(\bP) = \bQ$.
  \end{enumerate}
\end{proposition}

\Proof The implication (\ref{item:2}) $\Rightarrow$ (\ref{item:1}) is
trivial. Hence consider only the other direction. Here, the idea is to
define the map $\map{E}$ by 
\begin{equation} \label{eq:2}
  \map{E}(P_e) = Q_e \quad \forall e \in \sE.
\end{equation}
This map
is well defined because we have (by assumption) for each density
operator $\rho$ a second density operator $\sigma$ such that $\Tr[Q_e \rho] =
\Tr[P_e \sigma]$ holds for all $e \in \sE$. Hence if $\sum_e \lambda_e P_e = 0$ for
some real $\lambda_e$ we get 
\begin{equation}
  \sum_{e \in \sE} \lambda_e \Tr[\rho Q_e]=\sum_{e \in \sE} \lambda_e \Tr[\sigma P_e]=\Tr \left[\sigma \sum_{e \in \sE} \lambda_e P_e\right]=0.
\end{equation}
Since $\rho$ is arbitrary this implies $\sum_e \lambda_e Q_e = 0$. Therefore
$\map{E}$ defined in (\ref{eq:2}) is well defined, as stated. Using
the same reasoning we can show that $\map{E}$ is positive, which
concludes the proof. \qed \medskip

The difference between condition (\ref{item:2}) of this lemma and the
definition of $\gg$ in Equation (\ref{eq:3}) is the \emph{domain} of the
the map $\map{E}$. The following counter example which is taken (in a
slightly modified form) from \cite{paulsen} shows that such a map is in
general \emph{not extendible} as a positive map to $\Bnd{\sH}$.

Consider the diagonal $4\times4$ matrix $X = \diag(1,i,-1,-i)$ and the
operator system $\alg{S}$ spanned by $I,X,X^\dagger$. It is easy to
see that a general element $A = a I + b X + c X^\dagger$ is hermitian
iff $c = b^*$ and $a = a^*$ hold, and it is positive iff in addition
$a \geq 2\max( |\Re b|, |\Im b| )$ hence,
\begin{equation} \label{eq:6}
  A \geq 0 \Rightarrow c = b^*\ \text{and}\ a \geq \sqrt{2} |b|.
\end{equation}
Now consider the linear map 
\begin{equation} \label{eq:7}
  \alg{S} \ni A = a I + b X + c X^\dagger \mapsto \map{E}(A) =
  \left(\begin{array}{cc}
    a & \sqrt{2} b \\
    \sqrt{2} c & a
  \end{array} \right) \otimes I_2,
\end{equation}
where $I_2$ denotes the $2\times2$ unit matrix. Since a $2\times2$
matrix is positive iff its diagonal elements and its determinant are
positive, positivity of $\map{E}$ follows directly from Equation
(\ref{eq:6}). On the other hand we have $\|\map{E}(I)\| = 1$ and
$\|\map{E}(X)\|=\sqrt{2}$. Since $\|X\|=1$ this implies $\|\map{E}\|
\geq \sqrt{2} > \|\map{E}(I)\|$. But a positive map from a C* algebra
$\alg{A}$ into a a C* algebra $\alg{B}$ always satisfies \cite[Cor.
2.9]{paulsen} $\|\map{E}\| = \|\map{E}(I)\|$. Hence the map defined in
Equation (\ref{eq:7}) can not be extended to $\Bnd{\Bbb{C}^4}$ -- not
even to the abelian algebra generated by $I, X, X^\dagger$. As a
consequence of this reasoning we have shown that $\bP \Rip \bQ$
\emph{does not imply} $\bP \gg \bQ$.

Hence positive maps can in general not be extended as a
\emph{positive} map to a bigger algebra.  A very important special
case arises, however, if the map $\map{E}$ is \emph{completely}
positive. In this case a completely positive extension always exists
(cf. Theorem \ref{thm:arv}) This fact can be used along with
Proposition \ref{prop:1} to get an interesting characterisation of
$\succ$ in terms of ranges.

\begin{theorem}
  Consider two POVM's $\bP$, $\bQ$ with $|\bP| = |\bQ|$. Then the
  following statements are equivalent:
  \begin{enumerate}
  \item \label{item:3a} 
    $\bP \succ \bQ$
  \item \label{item:4} There is an informationally complete POVM $\bM$
    such that $\bP \otimes \bM \Rip \bQ \otimes \bM$.
  \item \label{item:5}
    $\bP \otimes \bM \Rip \bQ \otimes \bM$ holds for all POVMs $\bM$. 
  \end{enumerate}
\end{theorem}

\proof The implication (\ref{item:3a}) $\Rightarrow$ (\ref{item:5}) follows from the fact that 
(\ref{item:3a}) implies the existence of a map $\map{E}$ such that $\bQ=\map{E}(\bP)$, and trivially
the map $\map{E}\otimes\map{I}$ connects $\bP \otimes \bM$ with $\bQ \otimes \bM$, whence the
statement via Eq. (\ref{martellata}). Implication (\ref{item:3a}) $\Rightarrow$ (\ref{item:4}) is
just a special case of the previous one. Implication (\ref{item:5}) $\Rightarrow$ (\ref{item:4}) is
trivial. Hence only  (\ref{item:4})~$\Rightarrow$~(\ref{item:3a}) remains to be shown.

To this end note that informational completeness of $\bM$ implies  
\begin{equation}
  \Span(\bQ \otimes \bM) = \Span(\bQ) \otimes \Bnd{\sH},
\end{equation}
and similarly for $\bP \otimes \bM$. Therefore we have (according to
(\ref{item:4}) and Proposition \ref{prop:1}) a unique positive map 
\begin{equation}
  \map{F}: \Span(\bP) \otimes \Bnd{\sH} \to \Span(\bQ) \otimes \Bnd{\sH}
\end{equation}
with
\begin{equation} \label{eq:5}
  \map{F}(\bP \otimes \bM) = \bQ \otimes \bM.
\end{equation}
If we can show that $\map{F}$ has the form
\begin{equation} \label{eq:4}
   \map{F} = \map{E} \otimes \map{I}
\end{equation}
with a positive map $\map{E}: \Span(\bP) \to \Span(\bQ)$ and the
identity $\map{I}$ on $\Bnd{\sH}$, the theorem is proved because:
\begin{itemize}
\item Due to Equation (\ref{eq:4}) and positivity of $\map{F}$ the map
  $\map{E}$ is completely positive as a map on the \emph{operator
    system} $\Span(\bP)$. Hence by theorem \ref{thm:arv} it is
  extendible to a completely positive map on $\Bnd{\sH}$.
\item Equations (\ref{eq:5}) and (\ref{eq:4}) imply $\map{E}(\bP) =
  \bQ$ and therefore $\bP \succ \bQ$.
\end{itemize}

To prove Equation (\ref{eq:4}) firstly note that (\ref{item:4})
implies $\bP \Rip \bQ$. This follows from (with $e \in \sE(\bQ)$ and a
density matrix $\rho$ on $\sH$):
\begin{align}
  \Tr[\rho Q_e] &= \Tr \left[ \frac{\rho \otimes I}{d} \left(Q_e \otimes \sum_{f \in \sE(\bM)} M_f\right)
  \right] \\
  &= \sum_{f \in \sE(\bM)} \Tr\left[ (Q_e \otimes M_f) \left( \rho \otimes \frac{I}{d}
    \right) \right]
\end{align}
because we have by assumption a density matrix $\sigma$ on $\sH
\otimes \sH$ such that
\begin{equation}
   \Tr\left[ (\bQ \otimes \bM) \left( \rho \otimes \frac{I}{d} \right) \right] = \Tr
   \bigl[ (\bP \otimes \bM) \sigma\bigr]
\end{equation}
which in turn implies
\begin{align}
  \Tr[\rho Q_e] &= \sum_{f \in \sE(\bM)} \Tr\bigl[ (P_e \otimes M_f) \sigma\bigr] \\
  &= \Tr \left[ P_e \otimes \left( \sum_{f \in \sE(\bM)} M_f \right) \sigma \right] \\
  &= \Tr \bigl[ (P_e \otimes I) \sigma \bigr] = \Tr[ P_e \Tr_2 \sigma],
\end{align}
where $\Tr_2$ denotes the partial trace over the second tensor factor.
Hence $\Tr[\rho \bQ] = \Tr[(\Tr_2\sigma) \bP]$ which implies $\bP \Rip
\bQ$ as stated.

Now we can apply again Propostion \ref{prop:1} and get a positive map
$\map{E}: \Span(\bP) \to \Span(\bQ)$ satisfying $\map{E}(\bP) = \bQ$
and therefore $\map{E} \otimes \map{I} (\bP \otimes \bM) = \bQ \otimes
\bM$. Since $\map{F}$ is uniquely determined by Equation (\ref{eq:5})
this implies $\map{F} = \map{E} \otimes \map{I}$, which completes the
proof.  \qed \medskip

This theorem gives a clear geometric picture for the relation $\succ$
and it helps to understand the difference between $\succ$ and $\gg$:
if $\bP \gg \bQ$ holds we find for each \emph{separable state} $\rho$
on $\sH \otimes \sH$ a second separable state $\sigma$ such that
$\Tr[\bQ \otimes \bM \rho] = \Tr[\bP \otimes \bM \sigma]$. Hence, if
$\bP \succ \bQ$ does not hold (but $\bP \gg \bQ$) there must be an
\emph{entangled} state $\rho$ such that the probability vector
$\Tr[\bQ \otimes \bM \rho]$ can not be reproduced by any expectation
value of $\bP \otimes \bM$. This can be rephrased as follows: If $\bP
\gg \bQ$ holds but not $\bP \succ \bQ$ we can reproduce the
distribution of outcomes of $\bQ$ measurements on \emph{one system} by
appropriate $\bP$ measurements, but there is information about
entangled states which can be gained only by $\bQ$ and not by $\bP$.

A second special case of Proposition \ref{prop:1} arises, when $\bQ$
is abelian (i.e. all elements of the POVM commute). In this case the
map $\map{E}$ constructed in Proposition \ref{prop:1} is a map into an
abelian algebra and therefore completely positive. Hence we get

\begin{theorem} \label{thm:abel-ranges}
  Consider two POVMs $\bP, \bQ$ with $|\bP| = |\bQ|$ and assume that
  $\bQ$ is abelian. Then $\bP \Rip \bQ$ and $\bP \succ \bQ$ are equivalent. 
\end{theorem}

\proof According to Proposition \ref{prop:1} there is a positive map
$\map{E}$ from $\Span(\bP)$ into the abelian C* algebra $\alg{A}$
generated by the elements of $\bQ$. According to Theorem
\ref{thm:abel-pos} this map is completely positive and by Theorem
\ref{thm:arv} therefore extendible as a completely positive map to
$\Bnd{\sH}$. Hence $\bP \Rip \bQ$ implies $\bP \succ \bQ$. Since the other
implication is trivial the proof is completed. \qed \medskip

Note that a similar result does not hold if $\bP$ is abelian and $\bQ$
is not. The counter example given after Proposition \ref{prop:1}
applies even in this case.

The result from Theorem \ref{thm:abel-ranges} is very useful, in
particular because the range $\Rng(\bP)$ of an abelian POVM has a very
simple structure, which is completely characterized by the joint
eigenvalues of the elements of $\bP$. To see this, consider a joint
set of eigenvectors $\psi_\alpha, \alpha = 1, \ldots, d$ and
\begin{equation}
  P_e = \sum_{\alpha=1}^d \lambda_{e,\alpha} |\psi_\alpha\>\<\psi_\alpha| \quad \forall e \in \sE.
\end{equation}
The joint eigenvalues vectors
\begin{equation}
  \vec{\lambda}_\alpha = (\lambda_{e,\alpha})_{e \in \sE} \in \Reals^{|\bP|}
\end{equation}
form a set of probability vectors (in the case of joint degeneracies
of the elements of $\bP$ some of them may coincide) and for each
convex linear combination
\begin{equation}
  \vec{p} = \sum_{\alpha = 1}^d p_\alpha \vec{\lambda}_\alpha,\quad p_\alpha \geq 0,\quad \sum_\alpha p_\alpha =
  1
\end{equation}
we can find a density operator ($\rho = \sum_\alpha p_\alpha |\psi_\alpha\>\<\psi_\alpha|$ will do)
such that $\vec{p} = \Tr[\rho \bP]$ holds. Hence the \emph{convex hull}
of the $\vec{\lambda}_\alpha$ satisfies $\Conv(\vec{\lambda}_1, \ldots, \vec{\lambda}_d) \subset
\Rng(\bP)$. On the other hand we have for each density operator $\rho$:
\begin{equation}
  \Tr[\rho \bP] = \sum_{\alpha=1}^d \langle\psi_\alpha, \rho \psi_\alpha\rangle \vec{\lambda}_\alpha
\end{equation}
which implies $\Tr[\rho \bP] \in \Conv(\vec{\lambda}_1, \ldots, \vec{\lambda}_d)$. Hence we have
just shown:

\begin{proposition}
  The range $\Rng(\bP)$ of an abelian POVM $\bP$ coincides with the
  convex hull of the $\vec{\lambda}_1, \ldots, \vec{\lambda}_d$. 
\end{proposition}

The most simple example arises in the case of \emph{effects},
i.e. measurements with two outcomes. Obviously, each effect is abelian
and has the form $\bP = \{P, I - P\}$ with a positive operator $P \leq
I$. If $\mu_1, \ldots, \mu_d$ are the eigenvalues of $P$ given in decreasing
order we get $\vec{\lambda}_\alpha = (\mu_\alpha, 1 - \mu_\alpha)$. Hence all $\vec{\lambda}_\alpha \in
\Reals^2$ are located on the graph of the function $\Reals \ni x \mapsto 1 - x
\in \Reals$, and $\vec{\lambda}_1$ respectively $\vec{\lambda}_d$ are the outermost
points. This leads immediately to the following characterization of
the relation $\succ$ for effects:

\begin{theorem}\label{th:keyl}
  The effect $\bP$ is ``cleaner'' than the effect $\bQ$, i.~e.
  $\bP\succ\bQ$ iff
\begin{equation}
[\lambda_M(P),\lambda_m(P)]\supseteq [\lambda_M(Q),\lambda_m(Q)].
\label{eq:condeff}
\end{equation}
\end{theorem}

  
\begin{corollary}
  Given two effects $\bP$ and $\bQ$, then $\bP\simeq\bQ$ iff
  $\lambda_M(P)=\lambda_M(Q)$ and $\lambda_m(P)= \lambda_m(Q)$.
\end{corollary}

\section{Pre-processing: equivalence in dimension two}\label{s:pre2}

For dimension two the cleanness equivalence $\simeq$ and the unitary
equivalence $\simeq_U$ coincide.
\begin{theorem} For two-level systems $\bP\simeq\bQ$ iff
$\bP\simeq_U\bQ$.
\end{theorem} \Proof If all the elements of both POVM are trivial,
i.~e. $P_e=\alpha_e I$ and $Q_e=\beta_e I$, $\forall e$, then the
thesis follows easily. Therefore, we will focus on the nontrivial
case, in which there exists at least one element $P_i$ of $\bP$ (or
$Q_i$ of $\bQ$) that is nontrivial. Then, first, also $Q_i$ (or $P_i$)
is not proportional to the identity, since otherwise $P_i=\map F(Q_i)=\alpha_i
\map{F}(I)=\alpha_i I$, which contradicts the hypothesis. Second, by
Theorem \ref{th:maxminei} one has
\begin{align}
  &P_i=\lambda_M(i)|\phi_M^{(i)}\>\<\phi_M^{(i)}|+\lambda_m(i)|\phi_m^{(i)}\>\<\phi_m^{(i)}|\,,\\
  &Q_i=\lambda_M(i)|\psi_M^{(i)}\>\<\psi_M^{(i)}|+\lambda_m(i)|\psi_m^{(i)}\>\<\psi_m^{(i)}|\,.
\end{align}
Now, by hypothesis, there exist channels $\map E$ and $\map F$ such that $Q_i=\map E(P_i)$ and
$P_i=\map F(Q_i)$. Then, by linearity, 
\begin{equation}
Q_i=\lambda_M(i)\map E(|\phi_M^{(i)}\>\<\phi_M^{(i)}|)+\lambda_m(i){\map E}(|\phi_m^{(i)}\>\<\phi_m^{(i)}|)\,.
\end{equation}
We will now consider
\begin{equation}
\Tr[Q_i|\psi_M^{(i)}\>\<\psi_M^{(i)}|]=\lambda_M(i)=\Tr[P_i\transp{\map E}(|\psi_M^{(i)}\>\<\psi_M^{(i)}|)]\,,
\end{equation}
and this clearly implies that $\transp{\map
  E}(|\psi_M^{(i)}\>\<\psi_M^{(i)}|)=|\phi_M^{(i)}\>\<\phi_M^{(i)}|$.
Analogous arguments lead to the conclusion that $\transp{\map
  E}(|\psi_m^{(i)}\>\<\psi_m^{(i)}|)=|\phi_m^{(i)}\>\<\phi_m^{(i)}|$,
and moreover $\transp{\map
  F}(|\phi_M^{(i)}\>\<\phi_M^{(i)}|)=|\psi_M^{(i)}\>\<\psi_M^{(i)}|$
and $\transp{\map
  F}(|\phi_m^{(i)}\>\<\phi_m^{(i)}|)=|\psi_m^{(i)}\>\<\psi_m^{(i)}|$.
By collecting all the eigenstates of nondegenerate $P_i$'s and $Q_i$'s
in two sets, namely,
\begin{equation}
\begin{split}
  &\transp{\map E}:\{|\psi_M^{(i)}\>\<\psi_M^{(i)}|,|\psi_m^{(i)}\>\<\psi_m^{(i)}|\}_i\longmapsto\{|\phi_M^{(i)}\>\<\phi_M^{(i)}|,|\phi_m^{(i)}\>\<\phi_m^{(i)}|\}_i\\
  &\transp{\map F}:\{|\phi_M^{(i)}\>\<\phi_M^{(i)}|,|\phi_m^{(i)}\>\<\phi_m^{(i)}|\}_i\longmapsto\{|\psi_M^{(i)}\>\<\psi_M^{(i)}|,|\psi_m^{(i)}\>\<\psi_m^{(i)}|\}_i\;.\\
\end{split}
\end{equation}
and applying Theorem \ref{th:caves} it follows that there exists a
unitary $U$ such that $Q_i=UP_iU^\dag$ for all nontrivial $Q_i$'s.
Clearly, the same unitary transformation maps the trivial elements.
\qed

\section{Pre-processing: cleanness for number of outcomes $n\leq d$}\label{s:pre-under}

\begin{lemma}\label{lem:cleann}
  For fixed number of elements $n\leq d$ the POVM $\mathbf P$ is clean
  iff $\lambda_M(P_i)=1$ for all $i$. Such condition is also
  equivalent to $\lambda_m(P_i)=0$ with multiplicity at least $n-1$,
  and each vector which is eigenvector with unit eigenvalue for some
  element $P_j$ must belong to the kernel of all other POVM elements.
\end{lemma}
\Proof We first prove that the condition is also equivalent to
$\lambda_m(P_i)=0$ for all $i$. Indeed, consider a normalized
eigenvector $|u\>$ of $P_j$ with eigenvalue 1, and suppose by absurd
that some element $P_i$ has $\lambda_m(P_i)>0$. Then
\begin{equation}
\<u|u\>=\sum_k\<u|P_k|u\>=\<u|P_j|u\>+\<u|P_i|u\>+\sum_{k\neq i,j}\<u|P_k|u\> >1,\label{uu}
\end{equation}
and in order to have no contradiction one must have $\<u|P_i|u\>=0$,
namely $\lambda_m(P_i)=0$. Notice that Eq. (\ref{uu}) also implies
that $\<u|P_k|u\>=0$ for all $k\neq j$, namely the same eigenvector
$|u\>$ of $P_j$ is eigenvector of all $P_k$ for all $k\neq j$.
Moreover, since there must be at least $n$ vectors as $|u\>$, each
being eigenvector of a different element $P_j$ corresponding to unit
eigenvalue, and since any two vectors must be orthogonal (since for
some $j$ they are eigenvectors corresponding to different eigenvalues
of $P_j$), this means that the $0$ eigenvalue for each POVM element
must have multiplicity at least $n-1$, and all the eigenvectors of any
element with eigenvalue 1 are in the kernel of the remaining elements.

We now prove that the condition is sufficient. Suppose that a POVM
$\bQ$ exists such that $\bQ\succ\bP$. Then by Lemma \ref{lem:eigmap}
$\{0,1\}\subseteq\Spec(Q_i)$ for all $i$. We then need to prove that
in this case $\bP\simeq\bQ$. From now on we will denote by $|u\>_i^P$
an eigenvector of $P_i$ with eigenvalue 1 and by $|u\>_i^Q$ the same
for $Q_i$. The proof is constructive: consider the map with
Stinespring form $\map E(X)=V^\dag(I\otimes X)V$, where
\begin{equation}
V=\sum_i\sqrt{P_i}\otimes|u\>_i^Q\,,
\end{equation}
and the notation $T=O\otimes|u\>$ denotes the operator defined as $T|\psi\>=O|\psi\>\otimes|u\>$ for
all $|\psi\>\in\sH$. It is clear that $\map E(Q_i)=P_i$. Similarly, consider the map $\map
F(X)=W^\dag(I\otimes X)W$, where 
\begin{equation}
W=\sum_i\sqrt{Q_i}\otimes|u\>_i^P\,.
\end{equation}
This is such that $\map F(P_i)=Q_i$. We proved that POVM's $\mathbf P$
such that $\lambda_M(P_i)=1$ for all $i$ are clean. We will now prove
that it is also a necessary condition. Consider indeed a generic POVM
$\mathbf Q$ such that at least for one outcome $j$ $\lambda_M(Q_j)<1$.
Then one can consider any POVM $\bP$ with $\lambda_M(P_i)=1$ for all
$i$ and construct the isometry
\begin{equation}
W=\sum_i\sqrt{Q_i}\otimes|u\>_i^P\,.
\end{equation}
It is clear that the Stinespring form $W^\dag(I\otimes X)W$ defines a
channel $\map E$ such that $\map E(P_i)=Q_i$ for all $i$. Then
$\bP\succ\bQ$. Moreover, by hypothesis $\lambda_M(P_j)>\lambda_M(Q_j)$
and then it is impossible that $\bP\simeq\bQ$. \qed

An immediate corollary is the following
\begin{corollary}\label{cor:cleanobs}
The only clean elements with $n=d$ are the observables.
\end{corollary}
\Proof In Lemma \ref{lem:cleann} for $n=d$ the iff condition is
equivalent to have eigenvalue $0$ with multiplicity $d-1$ for each
POVM element, namely each POVM element is rank one, and they are
orthogonal. \qed \bigskip

Allowing mapping between POVM's with different number of outcomes,
the situation simplifies:
\begin{theorem}\label{th:cleand}
  For $n\leq d$ outcomes the set of clean POVM's coincides with the set
  of observables.
\end{theorem}
\Proof Consider a generic POVM $P_i$ with $i=1,\dots,n\leq d$. This
can be always regarded as the pre-processing of any desired observable
$\{|i\>\<i|\}_{i=1,\ldots,d}$. In fact, using the isometry from $\sH$
to $\sH^{\otimes 2}$
\begin{equation}
V=\sum_{i=1}^n \sqrt{P_i}\otimes|i\>,
\end{equation}
the following channel expressed in the Stinespring form
\begin{equation}
\map M(X)=V^\dag(I\otimes X) V
\label{eq:stin}
\end{equation}
gives
\begin{equation}
\map M(|i\>\<i|)=P_i,\qquad i=1,\ldots d.
\end{equation}
For a POVM with $n<d$ outcomes (strictly), notice that it is
equivalent to a POVM with $d$ outcomes and $d-n$ vanishing elements.
On the other hand, for $n<d$ there is no channel that can increase the
number of outcomes back to $d$, whence a POVM with $n<d$ outcomes
cannot be clean.  For $n=d$ Corollary \ref{cor:cleanobs} asserts that
the only clean POVM's are the observables. \qed \bigskip

\section{Pre-processing: ordering of informationally complete POVM's}\label{s:preinfo}

\begin{lemma}\label{lem:infocinfoc}
  If the POVM $\bQ $ is infocomplete then every $\bP$ such that $\bP
  \succ\bQ $ is infocomplete, too.
\end{lemma}
\Proof For $d^2$ outcomes POVM's, $\mathbf P$ and $\mathbf Q$ are
infocomplete iff their elements are linearly independent.  Suppose by absurd
that there exists a nonnull vector of $d^2$ coefficients $c_i$ such
that $\sum_{i=1}^{d^2}c_iP_i=0$, then also
\begin{equation}
\map E\left(\sum_{i=1}^{d^2}c_iP_i\right)=0=\sum_{i=1}^{d^2}c_iQ_i=0\,,
\end{equation}
which contradicts the hypothesis.

If the number of outcomes is greater than $d^2$, suppose
\begin{equation}
\map E(X)=0\,,
\end{equation}
for some $X\neq0$, namely $\map{E}$ would have non trivial kernel, in
which case $\Span(\bQ)\subseteq\Rng(\map{E})\subset\Bnd\sH$, which
contradicts the hypothesis that $\bQ=\map{E}(\bP)$ is infocomplete.
Then ${\map E}$ is invertible. Now, $\bP$ must be infocomplete,
otherwise the inverse of $\map E$ would not have full rank, which is
absurd. \qed \medskip

The above theorem is immediately extended to any linearly independent
POVM $\bQ $. More interestingly, for any infocomplete POVM $\mathbf P$
one can prove the following lemma
\begin{lemma}\label{lem:iceqic}
If the POVM $\bP $ is infocomplete then every ${\mathbf
    Q}$ such that $\bP \simeq\bQ $ is infocomplete, too.
\end{lemma}
\Proof It follows immediately from definition of $\simeq$ and
Lemma \ref{lem:infocinfoc}. \qed \medskip

On the other hand, each POVM that is equivalent to an infocomplete one, is also unitarily equivalent
to it, namely, more precisely
\begin{theorem}\label{th:infoeq}
  If $\mathbf P$ is an infocomplete POVM, then $\bP \simeq\bQ $ iff
  $\bP\simeq_U\bQ $.
\end{theorem}
\Proof Since the POVM's $\bP $ and $\bQ $ must be both infocomplete by
the previous lemma, then the maps $\map E$ and $\map F$ are uniquely
defined, and are the inverse of each other. Then, by Theorem
\ref{th:wigner} $\map E(X)=UXU^\dag$ for some unitary $U$.
\qed\medskip An alternative elegant proof of the above theorem would
be the following.  \Proof By hypothesis, there exist $\map E$ and
$\map F$ such that $\map E(\bP)=\bQ$ and $\map F(\bQ)=\bP$. This means
that $\map F\circ\map E$ stabilizes the algebra generated by $\bP$,
that is $\Span(\bP)\equiv\Bnd{H}$. On the other hand, the commutant of
an infocomplete POVM is only the identity, since $[P_i,X]=0$ for all
$i$ implies $[A,X]=\sum_ia_i[P_i,X]=0$ for all $A\in\Bnd{\sH}$. This
fact along with Lemma \ref{lem:lind} implies that $\map F\circ\map
E$ is the identical map. The thesis is then a straightforward
consequence of Theorem \ref{th:wigner}.  \qed\medskip

\begin{corollary}\label{l:invneg}
  For each non unitary invertible channel $\map{E}$ on $\Bnd{H}$ there
  exists at least a pure state $\psi\in\sH$ such that
  $\dual{\map{E}}{}^{-1}(|\psi\>\<\psi|)\not\geq0$.
\end{corollary}
\Proof Let us
consider an extremal POVM with $d^2$ rank-one elements
$\{|\alpha_i\>\<\alpha_i|\}$ ${i=1,\ldots,d^2}$ (according to Ref.
\cite{extrpovm} such a POVM always exists for any dimension $d$, and
it is necessarily informationally complete). Assuming $\map{E}$
invertible, let's consider $Q_i=\map{E}^{-1}(|\alpha_i\>\<\alpha_i|)$.
The POVM $|\alpha_i\>\<\alpha_i|$ is clean since it is rank-one.
However, since it is also infocomplete, then $Q_i$ cannot be itself a
POVM, otherwise according to Theorem \ref{th:infoeq} it would be
unitarily equivalent to $|\alpha_i\>\<\alpha_i|$.  Moreover, being
both $|\alpha_i\>\<\alpha_i|$ and $Q_i$ infocomplete, the map
$\map{E}$ would be univocally defined, whence itself unitary,
contrarily to the hypothesis. Then, $\{Q_i\}$ is not a POVM.  However,
since the map $\map{E}$ is a channel, whence $\map{E}^{-1}$ must be identity preserving, one has
$\sum_iQ_i=I$, then necessarily at least one element $Q_j$ cannot be positive, namely
there exists a vector $\psi\in\sH$ for which
\begin{equation}
\<\psi|Q_j|\psi\><0.
\end{equation}
This inequality can be rewritten as follows
\begin{equation}
\Tr[|\psi\>\<\psi|\map{E}^{-1}(|\alpha_j\>\<\alpha_j|)]=\Tr[\dual{\map{E}}{}^{-1}(|\psi\>\<\psi|)|
\alpha_j\>\<\alpha_j|]<0,
\end{equation}
namely $\dual{\map E}{}^{-1}(|\psi\>\<\psi|)$ is not positive. \qed \medskip

We have also the following interesting theorem.
\begin{theorem}
  Every channel $\map{F}$ which maps the set of states $\sS$
  surjectively on itself, i.~e.  such that $\map{F}(\sS)\equiv\sS$, is
  necessarily unitary.
\end{theorem}
\Proof First, suppose that $\map{F}$ is invertible, then $\map{F}$
must be unitary, otherwise $\map{F}^{-1}(\sS)=\sS$ would not be
possible by Lemma \ref{l:invneg}.  On the other hand, if $\map{F}$ is
not invertible, then its range must have dimension strictly smaller
than $d^2$. Now, consider a rank-one infocomplete POVM $\bP$ with
$|\bP|=d^2$.  Clearly, some POVM element cannot belong to $\map
F(\sS)$, and this proves that $\map{F}(\sS)\subset\sS$ strictly, since
such normalized POVM elements are just pure states. \qed \medskip

For qubits this theorem has the simple geometric interpretation that
the Bloch sphere transformed under $\map{F}^{-1}$ for any invertible
non unitary $\map{F}$ becomes an ellipsoid which contains elements
outside the Bloch sphere. \bigskip


By definition, and according to Theorem \ref{th:infoeq} an
infocomplete POVM $\bP$ is clean iff $\map{E}^{-1}(\bP)$ is not a POVM
for all invertible non unitary maps $\map{E}$. This means that as soon
as the set $\sS$ of states is transformed by $\map{E}^{-1}$, the POVM
is able to detect at least one of the points in
$\map{E}^{-1}(\sS)-\sS$, say $\map{E}^{-1}(|\psi\>\<\psi|)$, since the
``probability distribution'' corresponding to
$\map{E}^{-1}(|\psi\>\<\psi|)$ is no longer positive.

\section{Pre-processing: ordering of rank-one POVM's}\label{s:pre-rank}
Intuitively one thinks that a rank-one POVM is clean. This is actually
true, and it is more precisely stated by theorem \ref{t:rank1} in this
section. In order to prove it, we first need the following
\begin{lemma}\label{lem:rank1}
  If the POVM $\bQ$ is rank-one (i.~e. each element
  $Q_i$ can be written as $Q_i=|w_i\>\<w_i|$), then for any POVM $\bP$
  such that $\bP\succ\bQ$, also $\bP$ is rank-one, and
  $\Tr[P_i]=\Tr[Q_i]$, $\forall i$.
\end{lemma}
\Proof Consider the following normalized vectors
\begin{equation}
|\tilde w_i\>=\frac1{\sqrt N_i}|w_i\>\,,\quad Q_i=N_i|\tilde w_i\>\<\tilde w_i|\,,
\end{equation}
where $N_i=\Tr[Q_i]=\N{w_i}^2$, whence $\sum_i N_i=d$. Suppose $\bP
\succ\bQ $, and $\map E(\bP)=\bQ$. Then one can easily verify the
following identity
\begin{equation}
N_i=\Tr[Q_i|\tilde w_i\>\<\tilde w_i|]=\Tr[\map E(P_i)|\tilde w_i\>\<\tilde w_i|]=
\Tr[P_i\dual{\map E}(|\tilde w_i\>\<\tilde w_i|)]\,.
\label{eq:rank1}
\end{equation}
Now, by the CPT property of $\dual{\map E}$, $\dual{\map E}(|\tilde
w_i\>\<\tilde w_i|)$ is a state and clearly the last expression in Eq.
\eqref{eq:rank1} is less than or equal to the maximum eigenvalue
$\lambda_M(P_i)$ of $P_i$. We have than the following situation
\begin{equation}
N_i\leq\lambda_M(P_i)\leq\Tr[P_i]\,.
\label{eq:ineqr1}
\end{equation}
By the normalization and positivity of POVM's, we have that
$d=\sum_iN_i=\sum_i\Tr[P_i]$ and $N_i\geq0$, $\Tr[P_i]\geq0$. These
conditions along with Eq. \eqref{eq:ineqr1} imply
\begin{equation}
N_i\equiv\Tr[P_i]\quad\forall i\,,
\end{equation}
and this in turn implies $\lambda_M(P_i)=\Tr[P_i]$, namely $P_i$ is
rank-one. \qed 
\medskip

We will now prove the following theorem 
\begin{theorem}\label{t:rank1}
  If $\bQ$ is rank-one, then $\bP \succ\bQ $ iff $\bP\simeq_U\bQ $.
  Namely, rank-one POVM's are clean.
\end{theorem}
\Proof First, notice that by Lemma \ref{lem:rank1}, $\bP\succ\bQ $
implies that $\bP$ is rank one with $\Tr[P_i]=\Tr[Q_i]$, for all $i$.
We have then
\begin{align}
&P_i=|v_i\>\<v_i|=M_i|\tilde v_i\>\<\tilde v_i|\,,\quad\N{\tilde v_i}=1\,,\\
&Q_i=|w_i\>\<w_i|=M_i|\tilde w_i\>\<\tilde w_i|\,,\quad\N{\tilde w_i}=1\,,
\end{align}
where $M_i\equiv\Tr[P_i]=\Tr[Q_i]$, consistently with Lemma
\ref{lem:rank1}. Now, by hypothesis we have
\begin{equation}
M_i=\Tr[\map E(P_i)|\tilde w_i\>\<\tilde w_i|]=\Tr[P_i\dual{\map E}(|\tilde w_i\>\<\tilde w_i|)]=
M_i\Tr[|\tilde v_i\>\<\tilde v_i|\dual{\map E}(|\tilde w_i\>\<\tilde w_i|)]\,.
\end{equation}
As a consequence, necessarily $\Tr[|\tilde v_i\>\<\tilde
v_i|\dual{\map E}(|\tilde w_i\>\<\tilde w_i|)]=1$, and by CPT property
of $\dual{\map E}$ this implies $\dual{\map E}(|\tilde w_i\>\<\tilde
w_i|)\equiv|\tilde v_i\>\<\tilde v_i|$. Notice that since $\dual{\map
  E}(I)=\sum_iM_i\dual{\map E}(|\tilde w_i\>\<\tilde
w_i|)=\sum_iM_i|\tilde v_i\>\<\tilde v_i|=I$, then $\dual{\map E}$ and
$\map E$ are unital, namely both CPT and CPI. Then, by applying
Theorem \ref{th:caves} one has $\bP\simeq_U\bQ$. The converse is
trivial. $\blacksquare$\medskip
\section{Conclusions}\label{s:conclusion}
In this paper we have introduced the notion of {\em clean} POVM's, namely which are not irreversibly
connected to another POVM via a quantum channel. We used the adjective ``clean'' for such POVM's
in the sense that they are not affected by ``extrinsical'' quantum noise from the action of a
channel which is in principle avoidable. We have seen that, quite unexpectedly, the {\em
  cleanness} property is largely unrelated to the convex structure of POVM's, and there are clean
POVM's that are not extremal and extremal POVM's that are not clean.

The classification problem of POVM's cleanness turned out to be much harder than that of their
extremality, and in this paper we gave a complete classification of clean 
POVM's only for number $n$ of outcomes $n\leq d$ ($d$ dimension of the Hilbert space), whereas for
$n>d$ we gave a set of either necessary or sufficient conditions, and an iff condition for the case of
informationally complete POVM's for $n=d^2$. The difficulty for classifying the case 
$n>d$ reflects analogous difficulties in the theory of quantum measurements in assessing the maximal
POVM cardinality needed to attain the accessible information, cardinality whose lower bound has been
shown to be actually larger than $d$\cite{Shor,Fuchs}.

The novel issue of clean POVM's naturally opens new problems in the theory of quantum information
and quantum measurements. Besides the problem of the general classification of cleanness, it raises
the problem of characterizing all POVM's achievable from a given one via a quantum channel, or,
reversely, of all POVM's which can be evolved toward a given one via a quantum channel. These are
only initial steps toward a thorough analysis of the general problem of the partial ordering induced
by channels on the convex set of measurements, an issue which is not an academic mathematical
problem, but which is relevant for engineering new quantum measurements with minimal 
available resources.

\section*{Acknowledgments}

\par We are grateful to Madalin Guta for interesting discussions. 
This work has been co-founded by EC and Ministero Italiano
dell'Universit\`a e della Ricerca (MIUR) through the cosponsored
ATESIT project IST-2000-29681 and Cofinanziamento 2003.  P.P.
acknowledges support from the Istituto Nazionale di Fisica della
Materia under project PRA-2002-CLON. R. W. acknowledges hospitality of
the QUIT group and partial support from European Science Foundation.
G. M. D. also acknowledges partial support from the Multiple
Universities Research Initiative (MURI) program administered by the
U.S. Army Research Office under Grant No. DAAD1900-1-0177.

\vfil

\begin{thebibliography}{50}
\bibitem{history} For an earlier history of the subject and a list of
  main contributors the reader is addressed to the bibliography in
  Ref. \cite{helstrom}.
\end{thebibliography}
\end{document}